\documentclass{svproc}
\usepackage{url}
\usepackage{graphicx}
\usepackage{ccicons}                              

\newcommand{\RNum}[1]{%
  \textup{\uppercase\expandafter{\romannumeral#1}}%
}

\newcommand{\qu}[1]{``#1''}
\usepackage{blindtext}
\usepackage{graphicx}
\usepackage[T1]{fontenc}
\usepackage{amssymb}
\usepackage{graphicx}
\usepackage{amssymb}
\usepackage{tabularx}
\usepackage{ragged2e}
\usepackage{comment}
\usepackage{comment}
\usepackage[utf8]{inputenc}
\usepackage[T1]{fontenc}
\usepackage{lmodern} 
\usepackage{diagbox}

\usepackage{amsmath}
\usepackage{lscape} 
\usepackage[figuresright]{rotating}
\usepackage{subcaption}
\usepackage{multirow}
\usepackage{amsmath}
\usepackage[colorlinks=true,
            linkcolor=red,
            urlcolor=blue,
            citecolor=blue]{hyperref}
\usepackage{comment}

\begin{document}
\mainmatter              
\title{Gender Detection on Social Networks using Ensemble  Deep Learning}
\titlerunning{Gender Detection using  Deep Learning}
%
%

\author{Kamran Kowsari\inst{1,2} \and Mojtaba Heidarysafa\inst{1} \and  Tolu Odukoya\inst{4} \and Philip Potter \inst{4} \and  Laura E. Barnes\inst{1,3} \and Donald E. Brown \inst{1,3}}
\authorrunning{K. Kowsari et al.} 
%
%
\institute{Department  of  Systems \&  Information  Engineering,  University  of Virginia, Charlottesville, VA, USA
\and
University of California, Los Angeles~(UCLA), CA, USA
\and
School  of  Data  Science,  University  of  Virginia,  Charlottesville,  VA, USA
\and 
Department of Politics, University of Virginia, Charlottesville, VA, USA
\\~\\
$^*$ Co-corresponding authors:~\href{mailto:kk7nc@virginia.edu}{kk7nc@virginia.edu}}

\maketitle              

\begin{abstract}
Analyzing the ever-increasing volume of posts on social media sites such as Facebook and Twitter requires improved information processing methods for profiling authorship. Document classification is central to this task, but the performance of traditional supervised classifiers has degraded as the volume of social media has increased. This paper addresses this problem in the context of gender detection through ensemble classification that employs multi-model deep learning architectures to generate specialized understanding from different feature spaces.
\end{abstract}

\section{Introduction}
From 2012 to 2017 the average time spent on social networking has increased from 90 minutes to 135 minutes. Every second approximately 6,000 tweets appear on Twitter, which amounts to about 350,000 tweets per minute, 500 million tweets per day, and 200 billion tweets per year~\cite{venkataramana2016analysing}. This volume demands increasingly sophisticated approaches to author profiling and classification. 

Much of the recent work on automatic author-profiling based on classification techniques has utilized supervised learning techniques which include classification trees, Na\"{i}ve Bayes, support vector machines (SVM), neural nets, and ensemble methods. Classification trees and na\"{i}ve Bayes approaches to automatic author-profiling provide good interpretable data but tend to be less accurate than other methods during data analysis~\cite{kowsari2019text}. 

Scientists from different areas and disciplines have presented work comparable as our approach described in this research paper. We have organized this research paper into four main sections including social networks, feature extraction, categorization (classification) methods and techniques, and deep learning for classification.

\subsection{Social Networks:} Social networks are structures with nodes that represent people or entities within a social context, and whose edges represent interactions, influence, and collaboration between the entities~\cite{liben2007link}. They are dynamic platforms that change quickly, acquiring new nodes and edges that signify new interactions between the entities~\cite{liben2007link}. Social networks allows individuals maintain social relationships in any society, finding people with similar interests, outlooks, and causes strengthen these networks~\cite{mislove2007measurement}. 

In recent years, researchers have attempted to perfect gender detection in these contexts of these networks. Social networks typically allow members to determine what name, race, age, they associate with their profile, but this self-reporting increases the likelihood of false identities within the network~\cite{peersman2011predicting}. These false identities can reduce the quality of analyses of content, intended audience, or structure. 

Early work attempting to determining the gender of social network users relied primarily on text analysis and focused on psychological indicators of the gender of the author~\cite{heidarysafa2019women}. For example, Pennebaker and Graybeal 2001~\cite{Pennebaker2015} tried to examine if text analysis could reveal the personality traits and gender of several subjects. More recently, Peersman~\textit{et. al.} 2011~\cite{peersman2011predicting}, in a bid to identify and curb pedophiles who falsify their identities on social networks, using natural language processing on data from the Dutch platform Netlog, to detect gender.
\begin{figure*}[!t]
    \centering
    \includegraphics[width=\textwidth]{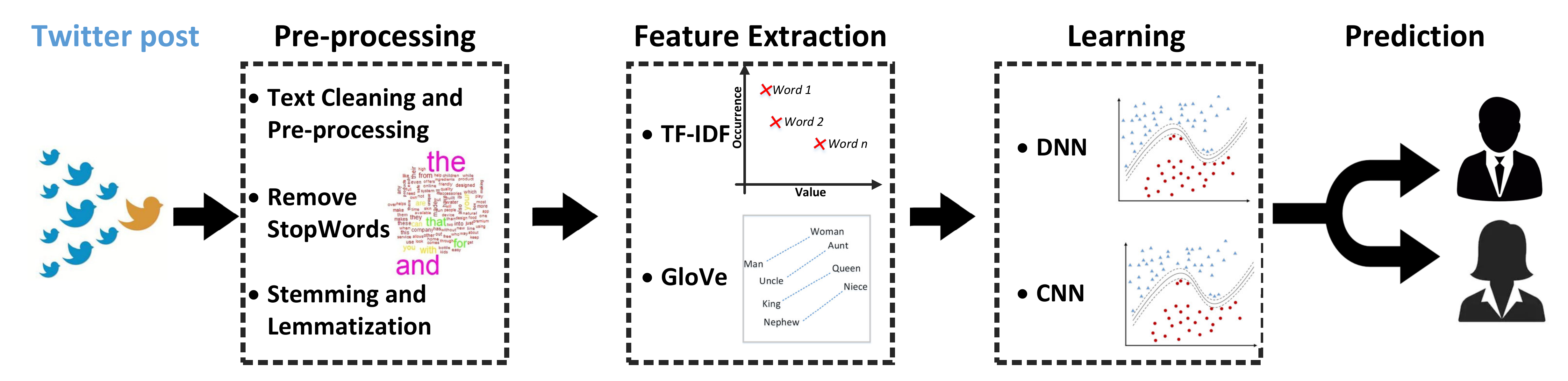}
    \caption{Pipeline of gender detection} \label{fig:pipeline}
\end{figure*}

\subsection{Feature Extraction:}\label{subsec:related1}
Feature extraction assumes a noticeable job Artificial intelligence, particularly for text, picture, and video information. Text and numerous biomedical datasets are generally unstructured information from which we have to create important structures for use by Machine Learning calculations. To take one early example, L. Krueger~\textit{et. al.} in~1979~\cite{krueger1979letter} presented a powerful technique for feature extraction dependent on word tallying to make a structure for statistical learning. Even prior work by H. Luhn~\cite{luhn1957statistical} presented weighted values for each word. In 1988 G. Salton~\textit{et. al.}~\cite{salton1988term} introduced the weights of words by frequency counts called term frequency-inverse document frequency~(TF-IDF) that  the vectors measure the occasions a word shows up in the datapoint (a document from a dataset), weighted by the opposite recurrence of the shared characteristic of the word across records. This technique~(word counting) is simple and intuitive feature extraction methods which they do not capture relationships between words as sequences.  T. Mikolov~\textit{et. al.}~\cite{mikolov2013efficient} presented a developed strategy for extracting features from documents utilizing the idea of installing or putting the word into a vector space dependent on context.  This technique, word embedding, called \textit{Word2Vec}, handles the issue of representing contextual word relationships in a commutable feature space. Building on these ideas, J. Pennington~\textit{et. al.} in~2014~\cite{pennington2014glove} built up a learning vector space portrayal of the words called~\textit{GloVe: Global Vectors for Word Representation} and deployed it in Natural Language Processing lab at Stanford University. The RMDL technique~\cite{Kowsari2018RMDL,Heidarysafa2018RMDL} uses two feature extraction methods including Glove and tf-idf from documents' dataset.

\subsection{Document Categorization Techniques and Methods:}\label{subsec:related2} 
Throughout the the past half-century, a numerous supervised learning methods as classification and categorization have been developed to accurately label data. For example, the scientist, K. Murphy in~2006~\cite{murphy2006naive} and I. Rish in~2001~\cite{rish2001empirical} present the Na\"ive Bayes Classifier~(NBC) as a straightforward way to deal with the more general representation of the supervised learning classification problem. This methodology has given a valuable strategy to text classification and information retrieval applications. Similarly, supervised learning classification techniques, NBC takes an input vector of numeric or categorical data values and produces the likelihood for every conceivable yield marks. This methodology is fast and efficient for text classification, but NBC has important limitations. Namely, the order of the sequences in the text is not reflected on the output probability because for text analysis, na\"ive bayes uses a bag of words approaches for feature extraction. Another well known classification method is Support Vector Machines~(SVM), which has proven quite accurate over a wide variety of data. This strategy develops a lot of hyper-planes in a changed element space. This change is not performed unequivocally but instead through the portion stunt which permits the SVM classifier to perform well with profoundly nonlinear connections between the indicator and reaction factors in the information. An assortment of approaches has been created to additionally expand the essential philosophy and acquire more prominent accuracy. C. Yu~\textit{et. al.} in 2009~\cite{yu2009learning} introduced latent variables into the discriminative model as a new structure for SVM, and S. Tong~\textit{et. al.} in 2001~\cite{tong2001support} added active learning using SVM for text classification. For an enormous volume of information and datasets with an immense number of features~(such as text), SVM executions are computationally complex. Another technique that helps mediate the computational complexity of the SVM for classification tasks is stochastic gradient descent classifier~(SGDClassifier)~\cite{kabir2015bangla} which has been extensively used in both document and visualized data categorization. This technique is an iterative method which mainly used for large datasets. SGD optimizer iteratively is modifying The weights till the model is trained.

\subsection{Deep Learning:}\label{subsec:related3}
Neural networks infer their design as a moderately basic representation of the neurons in the human's brain. They are basically weight mixes of data sources the go through numerous non-straight capacities. Neural networks utilize an iterative learning method known as back-propagation and an optimizer~(such as stochastic gradient descent~(SGD)). In recent studies, many scientists have achieved state-of-art results using Deep Learning in the domain of Social Media, Psychology~\cite{nobles2018identification}, transportation~\cite{heidarysafa2018analysis}, health~\cite{zhang2018patient2vec}, medical data processing~\cite{kowsari2019diagnosis}, etc.

Deep Neural Networks~(DNN) is based on simple neural network architectures which contains multiple hidden layers. Several researchers have used DNN for classification of text data. D. Cire{\c{s}}An~\textit{et. al.} in 2012~\cite{ci2012multitraffic} used multi-segment deep neural networks, which draws from DNN architectures, for classification tasks in their paper. Convolutional Neural Networks~(CNN) provide an alternative to the inherent DNN building structure to analyze learning within neural networks. CNN's main feature is the combining of feed-forward systems with convolutional builds that incorporate local and global pooling layers. A. Krizhevsky in 2012~\cite{krizhevsky2012imagenet} used CNN in their paper, however, their paper used a model that contained~$2D$ convolutional layers incorporated with ~$2D$ feature space of the analyzed image. Lecun groundbreaking work in 2015 ~\cite{lecun2015deep} used CNN to accomplish  excellent accuracy for image classification data. Kim successfully used Lecun's technique for~(CNN) to analyze text and document classification proving the versatility of Lecun's method \cite{kim2014convolutional}. For text, sequence, and document classification,~$1D$ convolutional layers use word embeddings as the input feature space. Novel research works combine different basic models of deep learning and develop new techniques for improving accuracy and robustness in classification tasks. M. Turan~\textit{et. al.} ~\cite{turan2017deep} and M. Liang~\textit{et. al.} ~\cite{liang2015recurrent} developed and implemented innovative combinations of RNN and CNN in their works called \textit{A Recurrent Convolutional Neural Network~(RCNN)}. K. Kowsari~\textit{et. al.} in 2017~ ~\cite{kowsari2017HDLTex} introduced hierarchical deep learning for text classification~(HDLTex) to the field. HDLTex is a technique that improves accuracy by combining deep learning techniques in a hierarchical structure for text classification.

%
%

\section{Preprocessing}\label{sec:preprocessing}
\subsection{Text cleaning}
\subsubsection{Tokenization}~\\
Tokenization is a part of the pre-processing technique that breaks the text of a document into words, phrases, symbols, or other meaningful elements called tokens~\cite{gupta2015text}.  The aim of tokenization is the investigation of the variety of words in a sentence~\cite{verma2014tokenization}. for example:\\
sentence~\cite{aggarwal2018machine} :

\textit{After walking for two hours, she decided to sleep.}\\
In this case, the tokens are as follows~\cite{kowsari2019text}:

\{ \textit{\qu{After} \qu{walking} \qu{for} \qu{two} \qu{hours} \qu{she} \qu{decided} \qu{to} \qu{sleep} } \}.

\subsubsection{Stop words}~\\
Text and document classification processes contain numerous words that are insignificant to the algorithm such as \{\textit{\qu{am},
\qu{is},
\qu{above},
\qu{are},
\qu{there},
\qu{his},
\qu{him},$\hdots$}\}. The most common method to solve this problem is to remove these words from the data~\cite{saif2014stopwords}.

\subsubsection{Capitalization}~\\
Text and document data contain several capitalized words to indicate the beginning of sentences within the corpus. Due to the volume of words in any corpus, diverse capitalization creates issues when classifying large documents. To resolve issues related to inconsistent capitalization every letter is reduced to lower case. The technique to resolve this issue, inputs all words in the document into the same feature space, but this solution can also create problems for the interpretation of certain words ~(i.e. "\textit{IT}"~(Information Technology) to "\textit{it}"~(subject pronoun))\cite{gupta2009survey}. Slang and abbreviation converters remedy the problem for these words \cite{dalal2011automatic}.

\subsubsection{Stemming and Lemmatization}~\\
Words can come in various forms~(i.e. the particular and plural thing structure) yet the semantic significance of each structure may remain the same~\cite{spirovski2018comparison}. Stemming is one strategy for solidifying various types of words into a similar component space. Text stemming changes words to get variation in the word structures by utilizing distinctive semantic procedures, for example, affixati (\textit{e.g.} the stem of "exampling" is "exampl"). Lemmatization, a NLP procedure, replaces the suffix with an alternate one or removes it entirely to get the essential word form~(lemma)

\subsection{Feature Extraction}

\subsubsection{Term Frequency-Inverse Document Frequency (TFIDF)}~\\ This technique is proposed inverse document frequency~(IDF)  which is a method that can be combined with term frequency to reduce the effects of implicitly common words in the dataset. 
IDF technique assigns higher weights to words that have either high or low frequencies in the document.
This TFIDF technique, a combination of Term Frequency and IDF, uses a mathematical equation (expressed below) to calculate the weight of a term in the data~\ref{tf-idf}.

\begin{equation}\label{tf-idf}
    W(d,t)=TF(d,t)* log(\frac{N}{df(t)})
\end{equation}

where~$N$ stands for the number of documents and~$df(t)$ indicates the number of documents containing the term~$t$ in the data. The first term in Equation~\ref{tf-idf} enhances the recall of the algorithm, and the second term enhances the accuracy of the word embedding~\cite{tokunaga1994text}. 

\subsubsection{Global Vectors for Word Representation~(GloVe)}~\\
The GloVe~\cite{pennington2014glove} approach is very similar to the Word2Vec technique that each word is assigned by high dimension vector which these vectors are trained based on the surrounding words. The pre-trained word embedding used in this work that is based on~$1,200,000$ vocabularies trained over~$2$-Billion tweets that contains~$27$ billion tokens, uncased, 25d, 50d, 100d, and 200d vectors. This word embedding is trained over even bigger corpora, including Wikipedia and Common Crawl content. The objective function is as follows:

\begin{equation}
    f(w_i-w_j,\widetilde{w}_k) =\frac{P_{ik}}{P_{jk}}
\end{equation}
where~$w_i$ refers to the word vector of any word~$i$, and~$P_{ik}$ as denoted, refers to the probability of any words~$k$ occurring in the same context of any word~$i$.

\begin{figure*}[t]
\centering
\includegraphics[width=\textwidth]{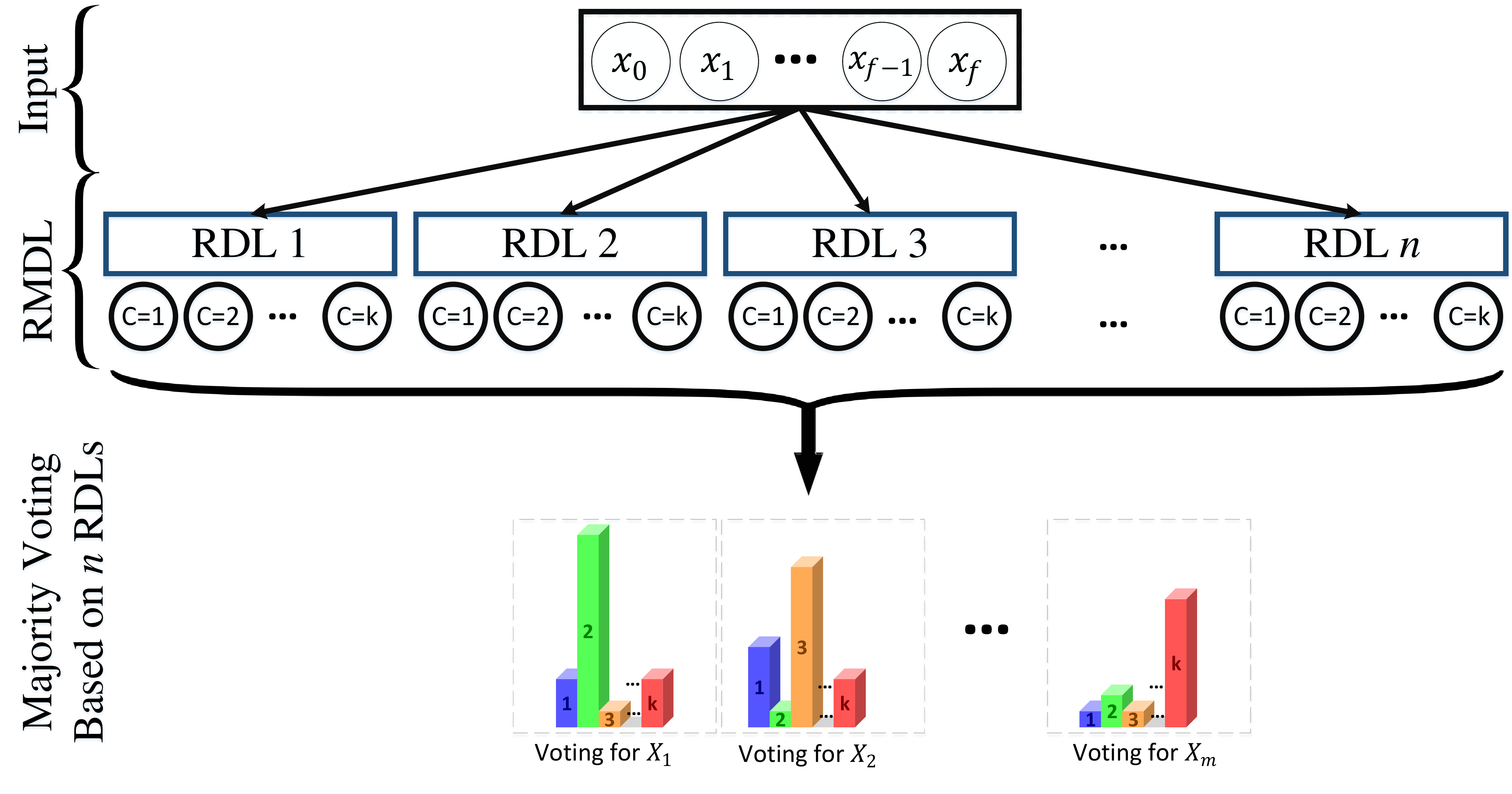}

\caption{RDML for gender detection which includes $n$ models which are $d$ random model of DNN classifiers, $c$ models of CNN classifiers where~$c+d=n$.}\label{Fig_RMDL}
\end{figure*}

\section{Methods}
The method used in this paper is based on Random Multimodel Deep Learning~(RMDL) for text and document categorization~\cite{Kowsari2018RMDL,Heidarysafa2018RMDL} in which we used two different feature extraction and ensemble deep learning algorithms to train this model. Random Multimodel Deep Learning is a new method that can be used in any kind of data classification task. The Figure~\ref{Fig_RMDL} shows how this algorithm works which contains multi Deep Neural Networks~(DNN) and Convolutional Neural Networks~(CNN). The structure and architecture of RMDL includes number of layers and nodes are generated randomly~(\textit{e.g.}~6 RDLs in  a RMDL constructed of~$3$ CNNs and~$3$ DNNs, all of them are usually unique due to randomly creation). 
\begin{align}
\label{eq:majority}
M(y_{i1},y_{i2},...,y_{in}) =& \bigg\lfloor \frac{1}{2}+ \frac{(\sum_{j=1}^n y_{ij}) - \frac{1}{2}}{n}\bigg\rfloor
\end{align}
where $n$ stands for the number of models, and $y_{ij}$ is the output of the model for data point $i$. Output space uses majority vote for final $\hat{y_i}$. Therefore,~$\hat{y_i}$ is given as follows:
\begin{equation}
\hat{y_i} =  
\begin{bmatrix}
\hat{y}_{i1} ~
\hdots ~
\hat{y}_{ij}~
\hdots~
\hat{y}_{in}~
\end{bmatrix}^T
\end{equation} 

where $n$ refers to number of  the models, and $\hat{y}_{ij}$ stands for the prediction of $D_i \in \{x_i,y_i\}$ for model $j$ and $\hat{y}_{i,j}$ is defined as follows:
\begin{equation}
\hat{y}_{i,j} = arg \max_{k} [ softmax(y_{i,j}^*)]
\end{equation}
After all models are trained, the majority vote is a final prediction of RMDL model. 

\begin{figure}[h]
    \centering
    \includegraphics[width=\textwidth]{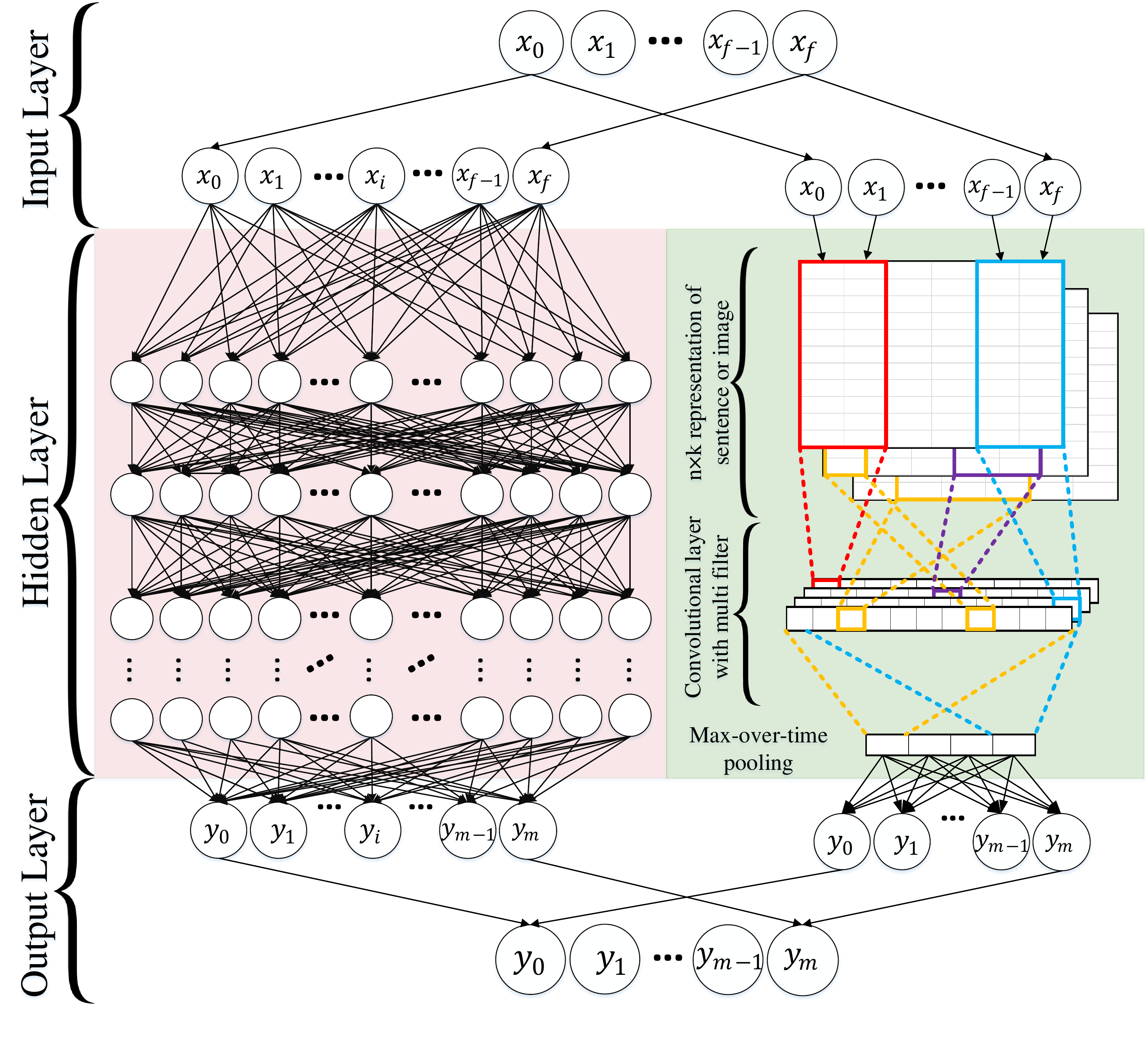}
    \caption{Random Multimodel Deep Learning~(RMDL)} \label{fig:RMDL}
\end{figure}

\subsection{Deep Neural Networks}
Deep neural networks~(DNN) are designed to learn by assuming a multi-layer-connection which states that every single layer only accepts the connection from the prior layer and supplies connections only to the following layer in the hidden part~\cite{kowsari2017HDLTex}. The input  comprises of the association of the input feature space with the first layer. The number of output layer is equal to the number of category for this research, we have only one node.

\subsection{Convolutional Neural Networks~(CNN)}\label{sec:CNN}
Convolutional Neural Network~(CNN) is a deep learning program that is often used in document classification tasks~\cite{jaderberg2016reading} hierarchy. Although it was in initially designed for image processing~\cite{kowsari2020hmic}, CNNs have also been shown to excel at  text classification~\cite{lecun2015deep}. In a basic CNN architecture, an image tensor is convoluted with a set of kernels of size $d \times d$. These newly merged layers known as feature maps, can be stacked to generate multiple filters on any input. To minimize the computational difficulty, CNNs use pooling to limit the size of any output from one layer in the network. A variety of  pooling techniques help minimize outputs while preserving important features~\cite{scherer2010evaluation}.

Max pooling, one of the foremost pooling techniques, makes sure the maximum value in the 2D or 1D pooling window is selected. These pooled outputs are flattened into one column and then fed from the stacked featured maps to the next layer. This usually results in the final layers of the CNN output being fully connected.
During the back-propagation process, the CNN feature detector filters and weights  are adjusted. However, this approach of using CNN for document categorization can create a problem since it can generate a large number of 'channels', $\Sigma$~(features). as shown in Figure~\ref{fig:RMDL}, the right box demonstrate the CNN structure for document categorization that include word embedding as input, and 1D convolutional layers followed by pooling layer (1D Maxpooling). The fully connected layers are connected to last flatten pooling layer, and finally output layer. 

\subsubsection{Adam Optimizer}

Adam, used as a stochastic gradient optimizer which utilizes just the initial first and second moments of gradient~(as shown in Equations~\eqref{adam}--\eqref{adam3}, $v$ and $m$ ) and calculates the mean. This optimizer can process nonstationary aspects of the objective function as in RMSProp while overcoming the sparse gradient issue limitation of RMSProp~\cite{kingma2014adam}.
\begin{align}
\theta  &\leftarrow \theta - \frac{\alpha}{\sqrt{\hat{v}}+\epsilon} \hat{m}\label{adam}\\
g_{i,t} &=  \nabla_\theta J(\theta_i , x_i,y_i) \label{adam1}\\
m_t &= \beta_1 m_{t-1} + (1-\beta_1)g_{i,t}\label{adam2}\\
m_t &= \beta_2 v_{t-1} + (1-\beta_2)g_{i,t}^2\label{adam3}
\end{align}
where $m_t$ is the first moment and $v_t$ indicates second moment that both are estimated. $\hat{m_t}=\frac{m_t}{1-\beta_1^t}$ and $\hat{v_t}=\frac{v_t}{1-\beta_2^t}$.

\begin{table*}[!t]
\centering
\caption{Method results}\label{tb:results}
\begin{tabular}{ c c l }
\hline
Measure                          & Value  & Derivations                                           \\ \hline
Sensitivity                      & 0.8914 & $TPR = \frac{TP}{(TP + FN)} $                           \\ \hline
Specificity                      & 0.8390 & $SPC = \frac{TN }{ (FP + TN)}$                                  \\ \hline
Precision                        & 0.8272 & $PPV = \frac{TP }{ (TP + FP)} $                               \\ \hline
Negative Predictive Value        & 0.8992 & $NPV = \frac{TN }{(TN + FN)}$                                  \\ \hline
False Positive Rate              & 0.1610 & $FPR = \frac{FP }{(FP + TN)}$                                  \\ \hline
False Discovery Rate             & 0.1725 & $FDR = \frac{FP }{(FP + TP)}$                                  \\ \hline
False Negative Rate              & 0.1086 & $FNR = \frac{FN }{(FN + TP)}$                                  \\ \hline
Accuracy                         & 0.8633 & $ACC = \frac{(TP + TN) }{(P + N)}$                             \\ \hline
F1 Score                         & 0.8583 & $F1 = \frac{2TP }{(2TP + FP + FN)}$                            \\ \hline
Matthews Correlation Coefficient & 0.7285 & $\frac{TP \times TN - FP \times FN }{\sqrt{ ((TP+FP)  (TP+FN)  (TN+FP)  (TN+FN) }}$ \\ \hline
\end{tabular}
\end{table*}

\section{Results}
\subsection{Data}
Demographic traits such as gender and language are historically investigated separately. PAN provided participants of its competition with a Twitter dataset annotated with all authors' gender and specific variations of their native languages: English (Great Britain, Ireland, Canada, New Zealand, Australia, and United States). In this research, we use combination of gender detection of Twitter posts~\cite{rangel2013overview,rangel2015overview} which contains large number of anonymous authors labeled with gender. This dataset is balanced by gender with 3600 people in the training set and  2400 for the test set.
\subsection{Evaluation and Experimental Results}\label{sec:Eva}
The underlying  mechanics of several assessment measurements may fluctuate, understanding what precisely every one of these measurements speaks to and what sort of data they are attempting to pass on is pivotal for equivalence. A few instances of these measurements incorporate  precision, recall, accuracy, F-measure, macro average and micro-average . These metrics are calculated from a~``confusion matrix'' that comprises true positives~(TP), false positives~(FP), false negatives~(FN) and true negatives~(TN)~\cite{lever2016points}. The importance of these four components may change dependent on the categorization algorithms. The division of right forecasts overall expectations is called accuracy~(Eq. \ref{eq:acc}). The fraction of known positives that are correctly predicted is called sensitivity~\textit{i.e.} true positive rate or recall~(Eq. \ref{eq:recall}). The ratio of correctly predicted negatives is called specificity~(Eq. \ref{eq:spec}). The proportion of correctly predicted positives to all positives is called precision,~\textit{i.e.} positive predictive value (Eq. \ref{eq:pres}). 

\begin{align}
    accuracy&=\frac{(TP+TN)}{(TP+FP+FN+TN)}\label{eq:acc}\\
    sensitivity&=\frac{TP}{(TP+FN)}\label{eq:recall}\\
    specificity&=\frac{TN}{(TN+FP)}\label{eq:spec}\\
    Precision &= \frac{\sum_{l=1}^LTP_l}{\sum_{l=1}^LTP_l+FP_l}\label{eq:pres}
\end{align}

$F_{\beta}$ is common techniques to aggregate evaluation metrics for any kind of classification evaluation~\cite{lever2016points}. The $F$ parameter of $\beta$ is used to balance recall and precision and defined as follows:

\begin{equation}{\label{eq:fbeta}}
F_{\beta} = \frac{(1+\beta^2)(precision \times recall)}{\beta^2 \times precision+recall}    
\end{equation}
For commonly used $\beta=1$~\textit{i.e.} $F_1$, recall and precision are given equal weights and Eq. \ref{eq:fbeta} can be simplified to:

\begin{equation}{\label{eq:f1}}
 F_1=\frac{2TP}{2TP+FP+FN}   
\end{equation}
Since $F_\beta$ is based on precision and , The confusion matrix cannot be regenerated from these measures.
The Matthews correlation coefficient~(MCC)~\cite{matthews1975comparison} catches all the information in a confusion matrix and measures the nature of paired order techniques (binary classification). MCC could be utilized for issues with unequal category sizes and could be considered a balanced score. The range of  MCC is from $-1$ to $0$ (i.e., the categorization can be always wrong and always true, respectively). The MCC is computed as follows:
\begin{equation}{\label{eq:mcc}}
MCC=\frac{TP \times TN - FP \times FN}{\sqrt{\frac{(TP+FP)\times(TP+FN)\times}{(TN+FP)\times(TN+FN)}}}
\end{equation}

Where analyzing binary classifiers, one may have a higher value using MCC and another one has a higher measure using $F_1$ and due to this specific metric fails to captures the benefits and drawbacks of a classifier~\cite{lever2016points}.

Figure~\ref{tb:results} shows our results by different measure as follows. Sensitivity of our model is~$0.8914$, and Specificity is equal to~$0.839$. The table shows Negative Predictive Value~(NPV) is~$0.8992$, False Positive Rate is~$0.161$. As we discussed the evaluation measures in this Section, Matthews Correlation Coefficient~(MCC) is~$0.7285$; and finally,  Accuracy and F1-Score is equal to~$0.8633$ and~$0.8583$ respectively.

\subsection{Hardware and Framework}
All of the works and experiment shown in this research paper are achieved on Graphical Process Units~(GPU) and Central Process Units~(CPU). The RMDL technique can be developed on only CPU, GPU, or both. The processing units that have been used through this explore was intel on \textit{Xeon~E5-2640~ (2.6 GHz)} with~\textit{12 cores} and~\textit{64~GB} memory. The GPU cards on the machine are~\textit{Nvidia Quadro~K620} and~\textit{Nvidia Tesla~K20c}. 

This RMDL for gender detection is is implemented by Python using CUDA~(Compute Unified Device Architecture) which is a parallel computing platform and API~(application programming interface) technique created by $Nvidia$. $Keras$ by using $TensorFelow$ back-end library is used for creating these deep learning algorithms~\cite{chollet2015keras}.

\section{Conclusion}
Developing methods for reliable gender detection using text classification is increasingly important the growing size of social text and other document sets. The techniques presented here demonstrate that that semantic, syntactic, and word frequency can facilitate gender detection in social messages. This approach improves on existing practice with different feature extraction techniques and deep learning architectures to train from a dataset as a ensemble learning technique. Additional training and testing with other structured text and message datasets will extend to identify architectures that work best for these problems. It is achievable to develop this model by using additional deep learning architecture as ensemble learning instead of two models to capture more of the complexity in the text classification.

\bibliographystyle{bibtex/spmpsci} 
\bibliography{template.bib}

\end{document}